# Skin Sympathetic Nerve Activity Driver Extraction through Non-Negative Sparse Decomposition

Farnoush Baghestani, Youngsun Kong, and Ki H. Chon, Fellow, IEEE
Biomedical Engineering Department, University of Connecticut, United States of America

*Abstract*—In recent years, skin sympathetic nerve activity (SKNA) extracted from electrocardiogram has gained attention as a novel noninvasive measure of the sympathetic nervous system (SNS), while electrodermal activity (EDA) has long served this purpose. SparsEDA is a sparse deconvolution technique originally developed for EDA to extract phasic drivers indicating the start of sympathetic burst responses. Our focus is on applying this method to preprocessed SKNA signals, justified by both SKNA and EDA signals' connection to sympathetic nerve activity and prior observed similarities. In a thermal-grill pain experiment, 16 subjects underwent six stimulations each to elicit SNS responses, with simultaneous recording of EDA and SKNA. We confirmed the method's accuracy in identifying stimuli initiation. Results were assessed for burst detection and accuracy of driver placement compared to annotated labels. The SKNA drivers achieved an RMSE of $0.42$ from annotated stimulations, a $97\%$ hit rate in detecting applied stimuli, and minimal false alarms ($1.40 \pm 1.76$) during the 2-minute control period and interstimulus intervals.

## I. Introduction

The study of the sympathetic nervous system (SNS) is of great importance, given its critical role in cardiovascular regulation and its established connection to conditions such as hypertension, coronary artery disease, and heart failure [1]. Recently, Doytchinova et al. proposed skin sympathetic nerve activity (SKNA) derived from human electrocardiogram (ECG) signals recorded at a sampling frequency of 2,000 Hz or higher as an effective SNS monitoring tool [2]. SKNA is extracted from frequency components ranging higher than the PQRST complex and myopotential signals ($< 500$ Hz). These frequency components were shown to stem from stellate ganglion nerve activity, which is associated with the SNS [3]. Subsequent studies also showed the feasibility of objectively assessing SNS activity using various SKNA indices [4]. Additionally, our previous study found that SKNA showed higher classification accuracy and faster stimulus-to-response time compared to electrodermal activity (EDA) [5].

Conventionally, the SNS has often been evaluated using EDA, which is a quantitative measure of sudomotor activity. EDA reflects the change in skin electrical conductance caused by the opening of sweat glands [6]. This change is closely linked to the SNS [6, 7]. Consequently, EDA has been a useful assessment tool for cognitive stress, fatigue, pain, sleepiness, depression, and other SNS-related applications [6]. EDA signals consist of two main components [8, 9]: 1) tonic and 2) phasic. The tonic component, also known as skin conductance level (SCL), represents the slow drifts and spontaneous fluctuations of the baseline skin conductance. The phasic component, known as the skin conductance response (SCR), exhibits fast dynamics in response to SNS stimuli. SCR exhibits a swift rise from the conductance level, followed by a gradual asymptotic exponential decay returning to the baseline.

To automatically decompose SCR and SCL and identify the phasic driver of SCR, various efforts have been made [8, 10, 11, 12, 13]. Among them, the SparsEDA algorithm [14] provides sparse phasic drivers for the SCR. This technique is computationally fast, and adaptable to any sampling rate and signal length for joint estimation of SCL and SCR components in the presence of additive white Gaussian noise. Modeling the SCR in SparsEDA builds upon a prior study [15] that involves the linear convolution between a non-negative sparse driver, representing sudomotor function (sympathetic activation), and the resulting response, actuated by the driver.

In our previous study, we found high correlations between a time series of SKNA, denoted as iSKNA, and its corresponding SCR during distinct Valsalva Maneuver stimulations, which involve an exaggerated breathing technique characterized by deep inhalation followed by a forceful and controlled expiration over time [5]. Therefore, we hypothesize that SparsEDA can be applied to iSKNA to obtain sparse SKNA drivers, because similar dynamics were exhibited between iSKNA and SCR bursts. In this study, we aim to extract suitable sparse drivers to mark SNS stimulations within SKNA signals using the SparsEDA approach. Additionally, given the tendency of iSKNA signals to exhibit subject-specific baselines, we also aimed to eliminate this individual variability. Identification of SKNA drivers can greatly facilitate analysis of SNS activity in experiments and applications without clear event onsets.

## II. Materials and Methods

### A. Participants and Experimental Setup

We recruited 16 healthy volunteers, consisting of 8 females and 8 males, aged 20 to 50, in the thermal-grill pain test [16]. The grills used in the study were constructed from interlaced copper tubes that circulated warm water (40–50°C) and cold water (18°C). This temperature contrast was designed to create an illusion of pain without causing tissue injury [16]. Within this setup, we employed two different grills to elicit low and high levels of pain stimulation. The distinction between the intensity of pain was disregarded, as our primary interest lies in the presence of stimulation and the temporal features. Therefore, we consider a total of 6 stimulations for each subject.

Blindfolded participants encountered the thermal grills on a wheeled table on their left. Using the verbal cue "ready," participants raised their left hand slightly from their lap, shifting it to the left. Once the intended grill was accurately positioned under the elevated hand of the subject, the conductor signaled with a "go" command. This cue instructed participants to promptly lower their left hand onto the grill. An interstimulus interval of about 40 seconds was conducted. Prior to stimulations, the subjects were given a 2-minute control period for recording, during which they were instructed to sit relaxed and minimize movements.

We captured the EDA signals from the index and middle fingers of the right hand. We also recorded an ECG channel. The sampling frequency was set at 10 kHz, via electrodes positioned on the inner wrists for Lead I configuration, and the ground electrode was placed under the left rib cage. GSR Amp and Bio Amp, connected to PowerLab and LabChart Pro 7 (ADInstrument, Sydney, Australia), were used to record EDA and ECG data, respectively. Subsequent processing was carried out using MATLAB R2022 (MathWorks, Natick, MA, USA). The study protocol was approved by the Institutional Review Board of the University of Connecticut.

*B. Preprocessing*

SKNA signals were derived using a bandpass filter from 1,700 Hz to 2,000 Hz instead of the conventional 500-1,000 Hz band. This adjustment, following the approach in [5], aimed to eliminate electrical noise artifacts from pump motors circulating water through the thermal grill tubes. Subsequently, the filtered signals were rectified by calculating the absolute value of the signal. iSKNA, representing the integral area of the rectified SKNA, was then calculated by applying a moving average filter with symmetric time windows of 100 ms, with 50 ms allocated for overlap at the start and end of each window.

The EDA signals were resampled to 4Hz. Prior to any analysis, through a visual inspection, we removed the segments corrupted by motion artifacts or other noise sources.

*C. Driver Extraction*

In this section, we provide a brief overview of the key components of the SparsEDA method [14] to support the rationale for its applicability to SKNA, a signal also considered to reflect skin nerve activity.

In SparsEDA, EDA was modeled to involve the linear convolution between a non-negative sparse driver, representing sudomotor function (sympathetic activation), $d_p(t)$, and the resulting response, $r(t)$, actuated by the driver, as follows:

$$s(t) = s_p(t) + s_l(t) = d_p(t) * r(t) + s_l(t), \quad (1)$$

where $s_p(t)$ and $s_l(t)$ represent phasic and tonic components, respectively.

SparsEDA employs a discrete-time version of equation (1):

$$s[n] = d_p[n] * r[n] + s_l[n] + w[n]. \quad (2)$$

In this expression, $s[n]$ signifies the EDA signal, where $d_p[n]$ is the sparse driver, $r[n]$ represents the resulting response, $s_l[n]$ corresponds to the SCL, and $w[n]$ denotes additive white Gaussian noise. Equation (3) represents the matrix notation of equation (2), where $R$ is an $N \times N$ (N being the length of the signal) Toeplitz matrix for $r[n]$ with $M$ non-zero values.

$$s = s_l + Rd_p + w. \quad (3)$$

This sparse recovery problem aims to jointly estimate $d_p[n]$ and $s_l[n]$ when $r[n]$ (and thus $R$) is unknown.

The authors of the SparsEDA method employed the modeling approach presented in [11] for sudomotor nerve activity, representing it as a biexponential function that characterizes the specific response triggered by the driver:

$$r(t) = e^{-\frac{t}{\tau_2}} - e^{-\frac{t}{\tau_1}}, \quad for\ t \geq 0 \quad (4)$$

where $\tau_2 = 0.75$ and $\tau_1 = 2$. This equation has been previously used to model the effects of individual nerve impulses on synaptic activation of the neuronal membrane [17]. However, the relationship between sudomotor bursts and the SCR is intricately influenced by the hydrodynamic properties of sweat pores. The SCR is more of an intricate process compared to the firing of a single synapse and entails a prolonged response. In contrast, SKNA represents a more direct measure of nerve activity.

With fixed values for $\tau_1$ and $\tau_2$, Hernando-Gallego et al. constructed waveforms with $Q$ different sampling periods during the discretization of $r(t)$ to address scale issues. The corresponding sparse vectors ($d_p$) form $d_{SCR}$:

$$R_{SCR} = [R_1, R_2, \ldots, R_Q] \quad (5)$$

$$d_{SCR} = [d_1^T, d_2^T, \ldots, d_Q^T]^T \quad (6)$$

A first order Taylor series expansion was used to approximate SCL:

$$s_l = [1/||1||_2 \ l/||l||_2 \ -l/||l||_2] \ d_{SCL} = R_{SCL}d_{SCL}, \quad (7)$$

where $1 = [1,\ldots,1]^T$, $l = [0,1,\ldots,N-1]^T$, and $d_{SCL}$ is a non-negative coefficient vector. By defining $R_T = [R_{SCL} \ R_{SCR}]$ and $d_T = [d_{SCL}^T \ d_{SCR}^T]^T$, the final formulation of the problem involves minimizing the mean squared error (MSE) between the available signal and its reconstructed counterpart:

$$\hat{d}_T = arg \min_{d_T} ||s - R_T d_T||_2^2 \quad (8a)$$

$$subject\ to\ d_T(i) \geq 0 \ \forall_i \quad (8b)$$

$$||d_T||_0 \ll NQ \quad (8c)$$

The sparsity constraint in Eq. (8c) is then relaxed by incorporating $L_1$ regularization term, leading to Least Absolute Shrinkage and Selection Operator (LASSO) [18]. The solution to LASSO is obtained through the LARS (Least-Angle Regression) algorithm [19]. However, as LARS is a greedy algorithm, the optimization process halts when $||s - R_T d_T||_2^2 \leq \epsilon$ or after a maximum number of iterations $K_{max}$ have been reached. The chosen values for termination conditions are consistent with those used by the original authors ($\epsilon = 10^{-4}$ and $K_{max} = 40$).

Finally, a postprocessing step enhances the sparsity of $d_p(t)$ by eliminating weaker impulses that are closely situated to stronger ones.

Subtracting the non-sparse output of the algorithm (SCR when the input signal is EDA) from iSKNA allows us to obtain a modified version of iSKNA devoid of the baseline.

*D. Statistical Analysis*

We examined intervals of 10 seconds in duration, commencing 1.5 seconds prior to each designated label indicating a stimulus, in order to address potential annotation errors. Within this specified valid window, if the algorithm assigned a driver, we consider it a "hit"; otherwise, the burst is considered a "miss." The Hit Rate for SKNA and EDA is calculated as the number of detected stimulations divided by the total number of applied stimuli, as expressed in Equation 9:

$$Hit\ Rate = \frac{Number\ of\ Detected\ Stimulations}{Total\ Number\ of\ Stimulations} \quad (9)$$

We proceeded to compute the stimulus-to-response time (delay) of the first-appearing driver within the window from the label. Additionally, the root mean squared error (RMSE) of the delay values from the labels was determined using Equation 10:

$$RMSE = \sqrt{\sum_{i=1}^{n} \frac{(\hat{y}_i - y_i)^2}{n}}, \quad (10)$$

where $n$, $\hat{y}$, and $y$ represent the total number of detected stimulations, the first-appearing driver, and the label, respectively. Any additional drivers identified outside the specified windows (baseline and interstimulus) were classified as false alarms.

## III. RESULTS

Fig. 1 provides an illustration of the decomposed signals and the corresponding drivers in a section of the pain experiment, including a portion of the control period followed by four pain stimulations, marked with vertical dotted lines at the onset of the stimulus. The top plot shows SCR along with its phasic drivers, marked by orange vertical lines, derived from the SparsEDA method to the EDA signal. The bottom plot depicts the output of the same method applied to the corresponding iSKNA signal. The baseline of the iSKNA signal is effectively removed by the subtraction of the non-sparse output. The SKNA drivers accurately mark the start of stimuli, with no false alarms in the control or interstimulus periods. In contrast, the last burst of EDA does not follow a driver (a missed burst). Additionally, the EDA drivers do not accurately mark the start of EDA bursts.

In general, the EDA bursts (hence phasic drivers) appear with a significant delay after the applied stimuli. Importantly, the first EDA driver after the first pain stimulation appears with a significant delay compared to the first SKNA driver. Considering all stimulations, while all SKNA drivers are located in a small window after the stimuli, there are usually additional EDA drivers in the following seconds, i.e. false alarms.

*A. Burst Detection*

Table 1 provides a summary of the burst detection results. Among all the valid EDA bursts, 77% were marked by phasic drivers within a valid window around the burst, whereas 97% of the SKNA bursts were successfully detected.

The second column in Table 1 summarizes the statistics of emergence of false alarms within the signals collected from the subjects. In an EDA signal, there were $12.39 \pm 5.44$ false alarms on average, whereas with SKNA signals, sparse decomposition generates only a few false alarms ($1.40 \pm 1.76$) in comparison.

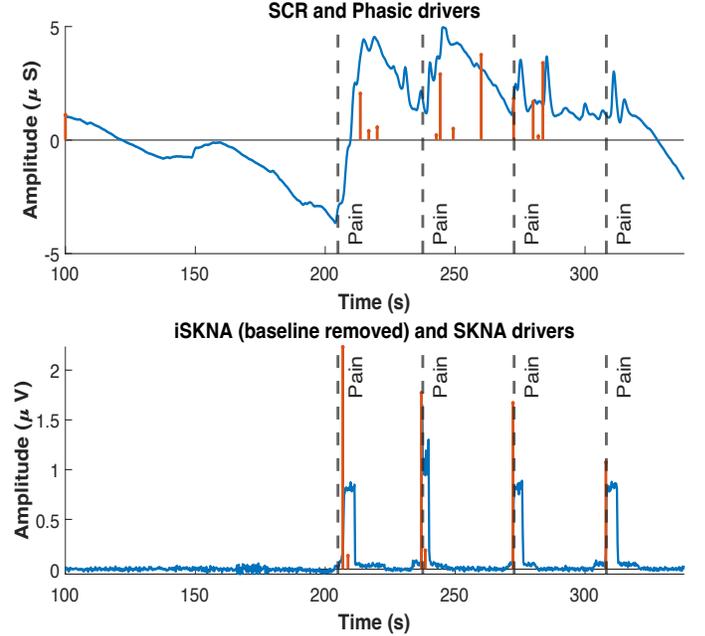

Fig. 1. Sparse Decomposition of EDA and iSKNA Signals. The vertical dotted lines indicate the application of sympathetic stimuli. The orange impulses show the calculated sparse drivers marking the stimuli.

TABLE I. BURST DETECTION RESULTS FROM SPARSE DRIVERS (MEAN ± S.D.)

|  | **Hit Rate** | **False Alarm [C.I]** |
|---|---|---|
| **EDA** | 77% | $12.93 \pm 5.44$ <br> $[9.79 - 16.07]$ |
| **SKNA** | 97% | $1.40 \pm 1.76$ <br> $[0.42, 2.38]$ |

*B. Temporal Features*

Fig. 2 illustrates the statistics (violin graphs) depicting the delays of EDA and SKNA drivers from the ground truth labels. We excluded missed bursts (with $delay \notin [-1.5, 10]$ s) from the calculation. Combining the information presented in the violin plots with that in Table 2, it becomes evident that, in general, SKNA drivers exhibit less variation and are more centered around the mean value, with an SKNA driver delay of $-0.12 \pm 0.40$ compared to an EDA driver delay of $4.73 \pm 2.53$. The negative values in the confidence interval of SKNA delay ($[-0.21, -0.04]$) indicate that the drivers typically occur before the annotated labels which are subjected to human labeling delays. The RMSE values for the drivers from the labels reveal that SKNA drivers (RMSE: 0.42) are significantly more accurate compared to EDA drivers (RMSE: 5.35).

TABLE II. SUMMARY OF THE TEMPORAL FEATURES COMPARED TO THE GROUND TRUTH LABELS (MEAN ± S.D.)

|  | Delay (s) [C.I] | RMSE |
|---|---|---|
| EDA | 4.73 ± 2.53 [4.09- 5.36] | 5.35 |
| SKNA | −0.12 ± 0.40 [−0.21- −0.04] | 0.42 |

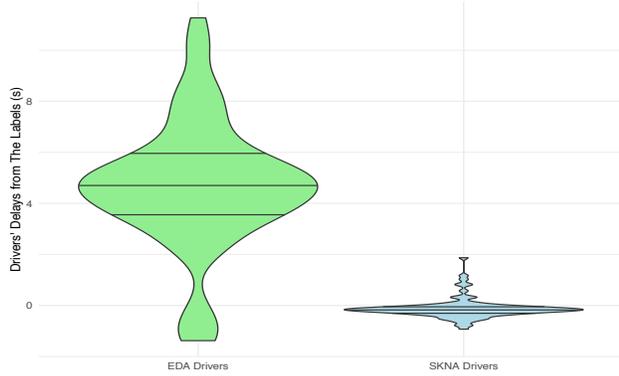

Fig. 2. Comparison of The Drivers' Delays

## IV. DISCUSSION AND FUTURE WORK

We evaluated a sparse decomposition method (SparsEDA), initially designed for EDA, to extract drivers for the SKNA signal. We compared its performance when applied to both SKNA and simultaneously recorded EDA signals. The decomposition method demonstrated effectiveness on iSKNA, providing precise representation of SKNA drivers with an RMSE of 0.42 seconds from annotated stimuli and achieving a 97% hit rate in detecting applied stimuli. In contrast, EDA drivers showed less favorable results, including an RMSE of 5.35 seconds and a lower detection rate of 77%. Additionally, there were fewer false alarms during the baseline and interstimulus intervals of iSKNA signals compared to those for EDA (1.40 ± 1.76 vs. 12.93 ± 5.44).

The majority of SKNA drivers preceded the annotations, indicating a swift response to sympathetic stimuli with a lead time of $−0.12 ± 0.40$ ($CI: [−0.21- −0.04]$) seconds. The emergence of SKNA drivers before the labels can be attributed to the potential for human error during label annotation and the time required for the conductor to press a key upon observing the subject's hand touching the grills. This lead time aligns with the human reaction time for visual stimuli which is around 0.27 seconds for a simple visual stimulus [20].

SparsEDA utilizes a biexponential function to model the response triggered by a phasic driver. The time constants used in this function were derived from a prior study that used a database of EDA signals from 735 subjects [11]. We acknowledge the necessity for future parameter optimization efforts using larger SKNA datasets to accommodate the abrupt and short nature of SKNA responses more effectively.

REFERENCES

[1] M. Sinski, J. Lewandowski, P. Abramczyk, K. Narkiewicz, and Z. Gaciong, "WHY STUDY SYMPATHETIC NERVOUS SYSTEM?".
[2] A. Doytchinova et al., "Simultaneous noninvasive recording of skin sympathetic nerve activity and electrocardiogram," *Heart rhythm*, vol. 14, no. 1, pp. 25–33, 2017.
[3] Z. Jiang et al., "Using Skin Sympathetic Nerve Activity to Estimate Stellate Ganglion Nerve Activity in Dogs," *Heart Rhythm*, vol. 12, no. 6, pp. 1324–1332, Jun. 2015, doi: 10.1016/j.hrthm.2015.02.012.
[4] Y. Xing et al., "An Artifact-Resistant Feature SKNAER for Quantifying the Burst of Skin Sympathetic Nerve Activity Signal," *Biosensors*, vol. 12, no. 5, Art. no. 5, May 2022, doi: 10.3390/bios12050355.
[5] Farnoush Baghestani, Youngsun Kong, William D'Angelo, Ki H. Chon, "Analysis of Sympathetic Responses to Cognitive Stress and Pain through Skin Sympathetic Nerve Activity and Electrodermal Activity," *Computers in Biology and Medicine*. Accepted.
[6] H. F. Posada-Quintero and K. H. Chon, "Innovations in electrodermal activity data collection and signal processing: A systematic review," *Sensors*, vol. 20, no. 2, p. 479, 2020.
[7] P. H. Ellaway, A. Kuppuswamy, A. Nicotra, and C. J. Mathias, "Sweat production and the sympathetic skin response: improving the clinical assessment of autonomic function," *Auton Neurosci*, vol. 155, no. 1–2, pp. 109–114, Jun. 2010, doi: 10.1016/j.autneu.2010.01.008.
[8] C. Setz, B. Arnrich, J. Schumm, R. La Marca, G. Tröster, and U. Ehlert, "Discriminating stress from cognitive load using a wearable EDA device," *IEEE Trans Inf Technol Biomed*, vol. 14, no. 2, pp. 410–417, Mar. 2010, doi: 10.1109/TITB.2009.2036164.
[9] A. Greco, G. Valenza, A. Lanata, E. P. Scilingo, and L. Citi, "cvxEDA: A Convex Optimization Approach to Electrodermal Activity Processing," *IEEE Transactions on Biomedical Engineering*, vol. 63, no. 4, pp. 797–804, Apr. 2016, doi: 10.1109/TBME.2015.2474131.
[10] W. Boucsein, *Electrodermal activity: Second edition*. 2013, p. 618. doi: 10.1007/978-1-4614-1126-0.
[11] D. M. Alexander, C. Trengove, P. Johnston, T. Cooper, J. P. August, and E. Gordon, "Separating individual skin conductance responses in a short interstimulus-interval paradigm," *Journal of neuroscience methods*, vol. 146, no. 1, pp. 116–123, 2005.
[12] M. Benedek and C. Kaernbach, "Decomposition of skin conductance data by means of nonnegative deconvolution," *Psychophysiology*, Mar. 2010, doi: 10.1111/j.1469-8986.2009.00972.x.
[13] M. Benedek and C. Kaernbach, "A continuous measure of phasic electrodermal activity," *Journal of neuroscience methods*, vol. 190, no. 1, pp. 80–91, 2010.
[14] F. Hernando-Gallego, D. Luengo, and A. Artés-Rodríguez, "Feature Extraction of Galvanic Skin Responses by Nonnegative Sparse Deconvolution," *IEEE Journal of Biomedical and Health Informatics*, vol. 22, no. 5, pp. 1385–1394, Sep. 2018, doi: 10.1109/JBHI.2017.2780252.
[15] C. L. Lim et al., "Decomposing skin conductance into tonic and phasic components," *International Journal of Psychophysiology*, vol. 25, no. 2, pp. 97–109, 1997.
[16] H. F. Posada-Quintero et al., "Using electrodermal activity to validate multilevel pain stimulation in healthy volunteers evoked by thermal grills," *American Journal of Physiology-Regulatory, Integrative and Comparative Physiology*, vol. 319, no. 3, pp. R366–R375, Sep. 2020, doi: 10.1152/ajpregu.00102.2020.
[17] J. J. Wright et al., "Toward an integrated continuum model of cerebral dynamics: the cerebral rhythms, synchronous oscillation and cortical stability," *Biosystems*, vol. 63, no. 1–3, pp. 71–88, 2001.
[18] R. Tibshirani, "Regression shrinkage and selection via the lasso," *Journal of the Royal Statistical Society Series B: Statistical Methodology*, vol. 58, no. 1, pp. 267–288, 1996.
[19] B. Efron, T. Hastie, I. Johnstone, and R. Tibshirani, "Least angle regression," *The Annals of Statistics*, vol. 32, no. 2, pp. 407–499, Apr. 2004, doi: 10.1214/009053604000000067.
[20] R. J. Kosinski, "A Literature Review on Reaction Time".